\documentclass[aip, jcp,groupedaddress,11pt,reprint]{revtex4-1}
\usepackage{amsmath}
\usepackage{color}
\usepackage{graphicx}
\renewcommand{\vec}[1]{\boldsymbol{#1}}
\begin{document}
\title{Tunneling-splittings from path-integral molecular dynamics using a Langevin thermostat}
\author{C L Vaillant}
\email{cv320@cam.ac.uk}
\author{D J Wales}
\author{S C Althorpe}
\email{sca10@cam.ac.uk}
\affiliation{Department of Chemistry, University of Cambridge, Lensfield Road, Cambridge CB2 1EW, United Kingdom}

\begin{abstract}
We report an improved method for the calculation of tunneling splittings between degenerate configurations in molecules and clusters using path-integral molecular dynamics (PIMD). Starting from an expression involving a ratio of thermodynamic density matrices at the bottom of the symmetric wells, we use thermodynamic integration with molecular dynamics simulations and a Langevin thermostat to compute the splittings stochastically. The thermodynamic integration is performed by sampling along the semiclassical instanton path, which provides an efficient reaction coordinate as well as being physically well-motivated. This approach allows us to carry out PIMD calculations of the multi-well tunnelling splitting pattern in water dimer, and to refine previous PIMD calculations for one-dimensional models and malonaldehyde. The large (acceptor) splitting in water dimer agrees to within 20$\%$ of benchmark variational results, and the smaller splittings are within $10\%$. 
 \end{abstract}

\maketitle

\section{Introduction}

Degenerate rearrangements of molecular clusters give rise to tunneling splittings in the rovibrational ground state. These splittings probe the potential energy surface sampled by the rearrangement dynamics. They are particularly useful in water clusters, for which the rearrangements sample a range of hydrogen-bonded structures, thus allowing universal water potentials\cite{Shank2009,Bukowski2007,Babin2013, Babin2014}  (designed to work in all phases of water) to be compared directly with experimental measurements of the splittings. The most studied water cluster, both experimentally\cite{Dyke1977,Zwart1991, Keutsch2003, Mukhopadhyay2015} and theoretically,\cite{Althorpe1991, Babin2013, Bukowski2007, Groenenboom2000, Chen1999, Wang2018, Leforestier1997,Gregory1995,Richardson2011b, Dyke1977, Hougen1985, Wales1996} is the dimer, which is now tractable by accurate variational methods including all 12 degrees of freedom.\cite{Wang2018} However, the larger clusters \cite{Liu1996, Gregory1996,Richardson2016, Wales1996} (for which the rearrangement dynamics sample three-dimensional hydrogen-bonded geometries) cannot be treated variationally.

There are a variety of non-variational methods for computing tunneling splittings.\cite{Miller1979,Gregory1995,Andersen1976, Bacic1989,Matyus2016a, Matyus2016b} Probably the simplest is the ring-polymer instanton (RPI) method,\cite{Richardson2011a, Richardson2011b,Cvitas2016, Cvitas2018, Milnikov2001, Milnikov2008, Siebrand1999} which has been applied to the water hexamer\cite{Richardson2016} and successfully reproduced experimental splitting patterns. However, the RPI method involves the steepest-descent approximation, and for this reason can produce errors if the fluctuations around the instanton path are too anharmonic. This situation arises for the facile acceptor-tunneling motion in the water dimer,\cite{Richardson2011b} and probably in the water hexamer, where the RPI splittings disagree with experiment by a factor of two.\cite{Richardson2016} To improve on semiclassical approximations, such as RPI, one must either make reduced-dimensionality approximations (which involve {\em a priori} assumptions about the dynamics) or use stochastic, imaginary-time-based approaches. The most established of these is fixed-node diffusion Monte Carlo (DMC) \cite{Andersen1976,Reynolds1982,Mizukami2014, Gregory1995, McCoy2012} which is often a highly practical method. However, DMC has the disadvantages that it requries prior knowledge of the nodal surface, and cannot converge very small splittings. Hence DMC cannot treat systems in which tunneling through a complex nodal surface gives rise to tiny splittings, as in the water hexamer prism ($\sim 10^{-5}$ cm$^{-1}$).\cite{Richardson2016}

For this reason, we developed in Ref.~\onlinecite{Matyus2016a} a path-integral molecular dynamics (PIMD) method for computing tunneling splittings, which builds on earlier path-integral Monte Carlo and molecular dynamics (MD) methods,\cite{Alexandrou1988, Marchi1991, Benjamin2005}  and is in the same class of path-integral methods as the RPI method. Instead of approximating the full path integral by steepest-descent, PIMD uses MD techniques to compute the integral stochastically. The advantage of PIMD is that the integrals involve the thermodynamic density matrix, rather than requiring separate calculations of ground and excited state energies. Thus relative splittings can be resolved without needing to converge individual absolute energy levels, permitting the calculation of much smaller splittings than DMC.

In Ref.~\onlinecite{Matyus2016a}, the PIMD method was tested on malonaldehyde, which has a large tunneling splitting ($\sim 21.6$ cm$^{-1}$), and for which the PIMD approach was 55 times slower than DMC, which seems likely to remain the most efficient method for calculating such large splittings. Here, we test the PIMD method on the water dimer, which is far less tractable by DMC (although early DMC calculations have been reported on this system\cite{Gregory1995}), having a splitting pattern consisting of a mixture of large and small splittings, produced by tunneling through complex nodal dividing surfaces. We also introduce two improvements to the method, namely a Langevin thermostat to facilitate sampling, and a thermodynamic integration path based on the instanton, which we compare with the previous implementation of PIMD in Ref.~\onlinecite{Matyus2016a}.

We begin by reviewing the theory of tunneling splittings using PIMD in section~\ref{theorysection}, before giving details of the new features of the method in section~\ref{newmethods}. Section~\ref{modelsection} reports the tests for one-dimensional models, malonaldehyde and water dimer. Section~\ref{discussionsection} concludes the article.

\section{Background Theory}
\label{theorysection}

\subsection{Tunneling Splittings from Symmetry-Related Density Matrices}
To calculate tunnelling splittings from path-integrals one needs an expression relating the splitting to the thermal density matrix. The standard approach, following Refs.~\onlinecite{Ceperley1995,Alexandrou1988, Marchi1991,Matyus2016a}, is to  take the density matrix in the position representation at two points related by a symmetry operation $\hat{P}$, such that $\psi_0 (\hat{P}\vec{r}) = \psi_0 (\vec{r})$ and $\psi_1 (\hat{P}\vec{r}) = -\psi_1 (\vec{r})$, leading to
\begin{subequations}
\begin{align}
\rho (\vec{r}, \hat{P}\vec{r}, \beta) &= | \psi_0 (\vec{r}) |^2 e^{-\beta E_0} - | \psi_1 (\vec{r}) |^2 e^{-\beta E_1} + \dots\\ 
\rho (\vec{r}, \vec{r}, \beta) &= | \psi_0 (\vec{r}) |^2 e^{-\beta E_0} + | \psi_1 (\vec{r}) |^2 e^{-\beta E_1} + \dots,
\end{align}
\label{densitymatrixsymmetry}%
\end{subequations}
where we take the low-temperature limit and include only the first excited state. The ratio between the two density matrices can then be written as
\begin{equation}
I(\beta)=\frac{\rho (\vec{r}, \hat{P}\vec{r}, \beta)}{\rho (\vec{r}, \vec{r}, \beta)} = \tanh \left[ \frac{\Delta}{2} (\beta - \bar{\beta})\right],
\label{tanhratio}
\end{equation}
where
\begin{equation}
\bar{\beta} = \frac{2}{\Delta} \frac{\psi_1(\vec{r})}{\psi_0(\vec{r})},
\label{betabar}
\end{equation}
is known as the tunneling time,\cite{Ceperley1995} and $\Delta$ is the tunneling splitting.

From here, one can calculate values of $I(\beta)$ for several $\beta$ and use least-squares fitting to obtain $\Delta$ and $\bar{\beta}$. Alternatively, one can calculate two values of $I(\beta)$ and solve \eqref{tanhratio} simultaneously, giving
\begin{subequations}
\begin{align}
\Delta &= 2 \frac{y^{(2)} - y^{(1)}}{\beta^{(2)} - \beta^{(1)}}\\
\bar{\beta} &= \frac{\beta^{(1)} y^{(2)} - \beta^{(2)} y^{(1)}}{y^{(2)} - y^{(1)}},
\end{align}
\label{simultaneoustanh}%
\end{subequations}
where $y^{(1)}$ and $y^{(2)}$ denote two values of $\tanh^{-1} [I]$ at $\beta=\beta^{(1)}$ and $\beta^{(2)}$, respectively.

If multiply degenerate wells are present in the system, then \eqref{tanhratio} needs to be generalized. The generalization is based on the use of a projection operator to extract the relevant contributions to the density matrix, as shown in Ref.~\onlinecite{Matyus2016b}. For large motion tunneling splittings, the wells and energy levels can be classified using permutation-inversion (PI) symmetry groups,\cite{LonguetHiggins1963, Bunker1998} where each well can be found through a sequence of ``feasible'' permutations of atoms (meaning permutations that do produce resolvable splittings) possibly combined with an inversion operation. The vibrational energy levels of the system can then be classified in terms of the irreducible representations (irreps) of the PI group, which are linear combinations of the PI operations of the group.

Consider a system with lowest energy wavefunctions $\psi_\nu (\vec{r})$ (where $\nu$ labels the irrep). We choose a basis such that the wavefunction for each level is described by a linear combination of wavefunctions centred about each well labelled $i$,
\begin{equation}
\psi_\nu (\vec{r})= \frac{1}{\sqrt{|G|}}\sum_i a_{\nu i} \phi (\vec{r}_i),
\label{eigenfunction}
\end{equation}
which is a consequence of applying the projection operator to the wavefunction centred on the initial well, $\phi(\vec{r})$.\cite{Bunker1998} Here, $\vec{r}_i = \hat{P}_i \vec{r}$, where $\hat{P}_i$ corresponds to the $i$th PI operation in the group of order $|G|$, and $a_{\nu i}$ is the character of the symmetry operation. A consequence of the assumption of symmetric wells implies that $\psi_\nu (\hat{P}_i \vec{r}) = a_{\nu i} \psi_\nu (\vec{r})/\sqrt{|G|}$. The density matrix for each well can then be written
\begin{equation}
\begin{split}
\rho (\vec{r}, \hat{P}_i \vec{r}, \beta) &= \sum_\nu \psi^*_{\nu} (\vec{r}) \psi_{\nu} (\hat{P}_i \vec{r}) e^{-\beta E_\nu}\\
&=\frac{1}{\sqrt{|G|}}\sum_\nu a_{\nu i} |\psi_{\nu} (\vec{r})|^2 e^{-\beta E_\nu},
\end{split}
\label{multiwelldensity}
\end{equation}
where $E_\nu$ is the energy of vibrational level $\nu$. We note that in \eqref{multiwelldensity} we make the low-temperature limit approximation and only sum over the ground and first-excited vibrational states.

Our goal is to obtain a similar expression to \eqref{tanhratio} that takes into account the multiwell nature of the density matrix. As shown in \eqref{multiwelldensity}, the density matrix connecting each well via its corresponding symmetry operation can be thought of as a linear combination of the contributions from different energy levels. Hence, we simply need to find the linear combination of density matrices required to isolate the contributions of individual energy levels. To achieve this, we make use of the orthogonality property of the characters,\cite{Bunker1998}
\begin{equation}
\frac{1}{|G|}\sum_i  a_{\nu'i} a_{\nu i} = \delta_{\nu' \nu},
\label{orthogonalitytheorem}
\end{equation}
which we combine with \eqref{multiwelldensity} to define new functions
\begin{equation}
\begin{split}
\eta^{\pm}_{\nu} (\beta)&= \frac{1}{\sqrt{|G|}}\sum_i \left[ a_{0 i} \pm  a_{\nu i}\right] \rho (\vec{r}, \hat{P}_i \vec{r}, \beta)\\
&= \frac{1}{\sqrt{|G|}}\sum_i \left[1 \pm  a_{\nu i}\right] \rho (\vec{r}, \hat{P}_i \vec{r}, \beta)\\
&=  |\psi_{0} (\vec{r})|^2 e^{-\beta E_0} \pm  |\psi_{\nu} (\vec{r})|^2 e^{-\beta E_\nu}.
\end{split}
\label{eta}
\end{equation}
We can follow the same procedure as \eqref{tanhratio} and define
\begin{equation}
I_\nu(\beta)=\frac{\eta^- (\beta)}{\eta^+ (\beta)} = \tanh \left[ \frac{\Delta_\nu}{2} (\beta - \bar{\beta}_\nu)\right],
\label{multiwelltanh}
\end{equation}
where we now have level-dependent values of $\Delta_\nu$ and $\bar{\beta}_\nu$, which are the multiwell analogues of $\Delta$ and $\bar{\beta}$ in \eqref{tanhratio}. Equations \eqref{simultaneoustanh} can then be used with $I_\nu$ calculated at two different values of $\beta$ to calculate the full set of tunneling splittings, with each calculation of $I_\nu$ requiring a calculation of $\rho(\vec{r}, \hat{P}_i \vec{r}, \beta)$ for each well.

\subsection{Path Integral Molecular Dynamics}
To calculate the value of $I(\beta)$ between two wells $\vec{a}$ and $\vec{b}$ using path integral methods, we first need to obtain expressions for a general density matrix. By $N$ insertions of the identity, we can rewrite a density matrix as
\begin{equation}
\rho (\vec{a}, \vec{b}, \beta) = \int \mathrm{d}\vec{r}_1 \dots \mathrm{d}\vec{r}_N \rho(\vec{a}, \vec{r}_1, \beta_N) \dots \rho(\vec{r}_N, \vec{b}, \beta_N),
\label{densitymatrix}
\end{equation}
which can be thought of as inserting multiple copies, or ``beads'', of the system between the two fixed points $\vec{a}$ and $\vec{b}$, where $\beta_N= \beta/(N+1)$. For small enough values of $\beta_N$, we can use the Trotter-Suzuki theorem to approximate the density matrices in \eqref{densitymatrix} as~\cite{Matyus2016a, Ceperley1995}
\begin{equation}
\rho(\vec{a}, \vec{b}, \beta) \approx \left( \frac{\prod_{j=1}^{f} m_j}{2\pi \hbar^2 \beta_N} \right)^{N/2} \int \mathrm{d}\vec{r}_1 \dots \mathrm{d}\vec{r}_N e^{-\beta_N U_N},
\label{approxdensity}
\end{equation}
where
\begin{equation}
\begin{split}
U_N &= \sum_{i=1}^{N} \left[ \sum_{j=1}^{f} \frac{1}{2} m_j \omega_N^2(r_{i,j} - r_{i+1,j})^2 +  V(\vec{r}_i)\right]\\
&+ \sum_{j=1}^{f}\frac{1}{2} m_j \omega_N^2 \left[ (a_{j} - r_{1,j})^2 + (r_{N,j} - b_{j} )^2 \right].
\end{split}
\label{un}
\end{equation}
Here, $f$ is the number of degrees of freedom (the number of atoms multiplied by the number of dimensions), $m_j$ is the mass of the atom for the $j$th degree of freedom of the $i$th bead, and $\omega_N = (\beta_N \hbar)^{-1}$.

The only practical way to evaluate the integrals in \eqref{approxdensity} is to use stochastic, Monte-Carlo-based methods. Here we use PIMD (with the thermostat to be described in Sec.~III.A). An effective classical Hamiltonian $H$ can be constructed from \eqref{densitymatrix} by inserting the identity, expressed as an integral over a Gaussian, $N \times f$ times to obtain
\begin{equation}
H = \sum_{i=1}^{f} \sum_{j=1}^N\frac{p_{i,j}^2}{2 \mu_{ij}} +  U_N.
\label{classicalhamiltonian}
\end{equation}
The ``bead mass'', $\mu_{ij}$, is a free parameter and is often chosen such that $\mu_{ij} = m_i$. We will return to this parameter later, explaining how a judicious choice of the bead mass can lead to improved convergence of molecular dynamics runs. Furthermore, we can collect together the terms relating only to the the polymer beads to construct an effective ``polymer Hamiltonian'',
\begin{equation}
\begin{split}
H_\mathrm{P} &= \sum_{i=1}^{N}\sum_{j=1}^f\left[  \frac{p_{i,j}^2}{2 \mu_{i,j}} + \frac{1}{2} m_j \omega_N^2(r_{i,j} - r_{i+1,j})^2 \right]\\
&+ \sum_{j=1}^{f}\frac{1}{2} m_j \omega_N^2 \left[ (a_{j} - r_{1,j})^2 + (r_{N,j} - b_{j} )^2 \right].
\end{split}
\label{polymerhamiltonian}
\end{equation}
The polymer Hamiltonian $H_\mathrm{P}$ can be expressed in polymer normal-mode coordinates to obtain a diagonal representation of $H_\mathrm{P}$. For the $k$th polymer normal-mode coordinate $q_{kj}$ acting on the $j$th degree of freedom with corresponding normal-mode frequency $\omega_{kj}$ (with $\beta_N^{-2}$ absorbed into the definition of our normal mode frequency), we can write $H_\mathrm{P} = \sum_{j=1}^{f} \sum_{k=1}^{N} H_\mathrm{P}^{(kj)}$, where
\begin{equation}
H_\mathrm{P}^{(kj)} = \frac{\pi_{kj}^2}{2 \mu_{kj}} + \frac{1}{2} m \omega_{k}^2 q_{kj}^2.
\label{polymermodeshamiltonian}
\end{equation}
Here, $\pi_{kj}$ is the conjugate momentum to $q_{kj}$. The normal-mode frequencies and the transformation between Cartesian and normal mode coordinates can be found in Ref. \onlinecite{Matyus2016a}.

\subsection{Thermodynamic Integration}

We employ thermodynamic integration to evaluate $I(\beta)$, which provides the free energy difference between two configurations by varying a tuning parameter interpolating between the configurations. We define a free energy function
\begin{equation}
F (\lambda, \beta_N) = -\frac{1}{\beta_N} \ln \rho(\vec{a}, \vec{b}(\lambda), \beta),
\label{freeenergy}
\end{equation}
where $\lambda$ is a reaction coordinate ($\lambda=0$, 1 for $\vec{r}=\vec{a}$, $\vec{b}$ respectively). From this definition, we can write
\begin{equation}
\begin{split}
I(\beta) &= \exp\left[ -\beta_N ( F(\lambda=1, \beta_N) - F(\lambda=0, \beta_N))\right]\\
&= \exp\left[ -\beta_N \Delta F \right],
\end{split}
\label{freeenergydifference}
\end{equation}
where the free energy difference $\Delta F$ is now the quantity to be calculated by dragging the fixed end bead from one well to another.

Following the standard thermodynamic integration technique, we write
\begin{equation}
\begin{split}
\Delta F &= \int_0^1 \mathrm{d} \lambda' \left. \frac{\partial F}{\partial \lambda} \right\rvert_{\lambda '}\\
&= \int_0^1 \mathrm{d} \lambda' \left\langle \frac{\partial H} {\partial \lambda} \right \rangle_{\lambda '},
\end{split}
\label{thermointegral}
\end{equation}
where $\langle \dots \rangle_{\lambda}$ denotes a thermodynamic ensemble average at a fixed value of $\lambda$. The derivative of the Hamiltonian can be performed analytically, resulting in~\cite{Matyus2016a}
\begin{equation}
\left\langle \frac{\partial H} {\partial \lambda} \right \rangle_{\lambda} = -\sum_{i=1}^{f} m_i \omega_N^2 \left\langle r_{iN} \right\rangle_{\lambda} \frac{\partial b_i}{\partial \lambda}
\label{hamiltonianderivative}
\end{equation}
Hence for a given value of $\lambda$ the ensemble average of the Hamiltonian derivative is simply the ensemble average of the position of the last bead before the fixed end bead. The integration in \eqref{thermointegral} is then achieved using Gauss-Legendre integration,\cite{Press1992} which allows us to perform the integral using a small number of $\lambda$ values.

\section{New Methodology}
\label{newmethods}
\subsection{Molecular Dynamics Simulations and Thermostats}
Path-integral MD trajectories are not in general ergodic,\cite{Hall1984} and thus need to be combined with a thermostat. The Andersen thermostat was used in our earlier PIMD calculations,\cite{Matyus2016a} but more efficient thermostats are available.\cite{Ceriotti2010} In this article, we will compare the local Andersen and Langevin thermostats (where the thermostats act on each degree of freedom independently). The Langevin thermostat has previously been used in PIMD simulations of the water-parahydrogen dimer,\cite{Constable2013, Schmidt2014} where the ground-state energy and wavefunction were efficiently calculated.

For the two thermostats, we use a velocity-Verlet integration scheme including the analytically-known polymer normal mode part of the propagation,\cite{Ceriotti2010} hence speeding up the convergence substantially. To further increase the efficiency, we combine the first few with the last few steps of the propagation so that the potential calculation only has to be performed once, thus halving the number of total calls to the potential. As an added bonus, this procedure also halves the number of thermostat steps (and thus the number of random numbers that need to be generated). The steps in the general integration scheme are therefore
\begin{widetext}
\begin{subequations}
\begin{align}
p_{ik} &\leftarrow p_{ik}  - \Delta t\frac{\partial V}{\partial r_{ik}}\\
\begin{pmatrix} \pi_{jk}\\ q_{jk}\end{pmatrix} &\leftarrow
\hat{U}_\mathrm{nm} \begin{pmatrix} p_{ik}\\ r_{ik} \end{pmatrix}\\
\begin{pmatrix} \pi_{jk}\\ q_{jk}\end{pmatrix} &\leftarrow
\begin{pmatrix} \cos{( \tilde{\omega}_{jk} \Delta t)} & -\tilde{\omega}_{jk} \mu_{jk} \cos{( \tilde{\omega}_{jk} \Delta t)}\\
\sin{( \tilde{\omega}_{jk} \Delta t)}/(\tilde{\omega}_{jk} \mu_{jk}) & \sin{( \tilde{\omega}_{jk} \Delta t)} \end{pmatrix}
\begin{pmatrix} \pi_{jk}\\ q_{jk}\end{pmatrix}\\
\begin{pmatrix} p_{ik}\\ r_{ik} \end{pmatrix} &\leftarrow
\hat{U}^{-1}_\mathrm{nm} \begin{pmatrix} \pi_{jk}\\ q_{jk}\end{pmatrix},
\end{align}
\label{velocityverlet}%
\end{subequations}
\end{widetext}
where $U_\mathrm{nm}$ denotes the linear polymer normal mode transformation,\cite{Matyus2016a} $i$ labels the bead, $j$ labels the degree of freedom and $k$ labels the polymer mode.
The frequencies are given by
\begin{equation}
\begin{split}
\tilde{\omega}_{jk} &= \omega_{k}\sqrt{\frac{m_j}{\mu_{jk}}}\\
&=\frac{2}{\beta_N \hbar }\sin{\left(\frac{k \pi}{2N+1}\right)}\sqrt{\frac{m_j}{\mu_{jk}}},
\end{split}
\label{polymerfreq}
\end{equation}
where $\omega_k$ denotes the polymer mode frequency.

 The Andersen thermostat is very simple to apply:\cite{Andersen1980} after a certain number of steps, one resets the normal mode momenta, $\vec{\pi}$, to a Gaussian-distributed random number with standard deviation $\sqrt{\mu_{jk}/\beta_N}$ and transforms back to the classical momenta $\vec{p}$. This procedure simulates periodic collisions with a fictitious bath of particles, ensuring a Boltzmann distribution of energy for the polymer.

The Langevin thermostat adds a stochastic drag term to the equations of motion that allows correct sampling of the canonical distribution.\cite{Ceriotti2010, Bussi2007} In our case, the polymer Hamiltonian in \eqref{polymermodeshamiltonian} is exactly a harmonic oscillator, and can be propagated analytically in the normal mode basis. Each degree of freedom in \eqref{polymerhamiltonian} is uncoupled and the overall system can be treated as separate 1D polymers to be summed over. Hence, for a Langevin diffusion process, Hamilton's equations for each 1D polymer are expressed in normal mode coordinates as~\cite{Gardiner1985, Bussi2007}
\begin{subequations}
\begin{align}
\frac{\mathrm{d} q_{jk}}{\mathrm{d}t} &= \frac{\pi_{jk}}{\mu_{jk}}\\
\frac{\mathrm{d} \pi_{jk}}{\mathrm{d}t} &= m_j \omega_k^2 q_{jk} - \sum_i \hat{U}_\mathrm{nm} V'(r_i)\nonumber\\&\quad - \gamma_{jk} \pi_{jk} + \sqrt{\frac{2 m_j \gamma_{jk}}{\beta_N}} \xi(t),
\end{align}
\label{langevin}%
\end{subequations}
where $\gamma_{jk}$ is the drag factor and $\xi(t)$ is a Gaussian-distributed random force. The noise term obeys $\langle \xi \rangle =0$ and $\langle \xi(t) \xi(0) \rangle = \delta(t)$ (here the angular brackets without a subscript denote a time average).

Hamilton's equations in \eqref{langevin} correspond to having an extra term in the Liouvillian
\begin{equation}
\mathcal{L}_\gamma = -\gamma_{jk} \left(\frac{\partial}{\partial \pi_{jk}} + \frac{m_j}{\beta_N} \frac{\partial^2}{\partial \pi_{jk}^2} \right).
\label{langevinliouvillian}
\end{equation}
If we use a symmetric Trotter splitting as before, the overall effect of the Langevin thermostat is to replace the first step of \eqref{velocityverlet} with~\cite{Ceriotti2010, Bussi2007}
\begin{subequations}
\begin{align}
p_{ik} &\leftarrow p_{ik}  - \frac{\Delta t}{2}\frac{\partial V}{\partial r_{ik}}\\
\pi_{jk} &\leftarrow \hat{U}_\mathrm{nm}  p_{ik} \\
\pi_{jk} &\leftarrow  e^{-\gamma_{jk} \Delta t/2} \pi^{(k)}_{jk} (t) + \sqrt{\left(1 - e^{-\gamma_{jk} \Delta t}\right)\frac{m_j}{\beta_N}} R\\
p_{ik} &\leftarrow \hat{U}^{-1}_\mathrm{nm}\pi_{jk}\\
p_{ik} &\leftarrow p_{ik}  - \frac{\Delta t}{2} \frac{\partial V}{\partial r_{ik}},
\end{align}
\label{langevinthermostat}%
\end{subequations}
where $R$ is a normally distributed random number with mean 0 and variance 1. The calculation of the potential still need only be performed once, as the positions $\vec{r}$ do not change in \eqref{langevinthermostat}.

We turn to the question of what value to assign the drag coefficients $\gamma_{jk}$. The ideal drag coefficient is the coefficient that minimizes the autocorrelation time.\cite{Ceriotti2010} However, for a general potential the contribution of $V$ in \eqref{langevin} typically cannot be found analytically. Nevertheless, we can find a good initial approximation using the polymer Hamiltonian autocorrelation time
\begin{equation}
\tau_H = \frac{1}{\langle H_\mathrm{P}^2\rangle - \langle H_\mathrm{P} \rangle^2} \int^\infty_0 \mathrm{d}t \langle \delta H_\mathrm{P}(0) \delta H_\mathrm{P}(t) \rangle,
\label{autocorrelationtime}
\end{equation}
where $\langle \delta H(0) \delta H(t) \rangle$ denotes the fluctuation correlation function. As shown in Appendix A, the autocorrelation time can be derived analytically for a damped harmonic oscillator with stochastic term, resulting in
\begin{equation}
\tau_H = \frac{\mu_{jk} \gamma}{4 m_j \omega_k^2} + \frac{1}{\gamma}.
\label{finalautocorrelation}
\end{equation}
The minimum autocorrelation time for the polymer Hamiltonian is therefore given by the friction coefficient
\begin{equation}
\gamma_\mathrm{P} = 2\omega_k \sqrt{\frac{m_j}{\mu_{jk}}}.
\label{polymerfriction}
\end{equation}
We modify the above expression to account for the potential by introducing an extra factor $\gamma_0$ to obtain a final friction coefficient
\begin{equation}
\gamma_{jk}= 2\gamma_0 \omega_k \sqrt{\frac{m_j}{\mu_{jk}}}.
\label{friction}
\end{equation}
The $\gamma_0$ factor is a free parameter that can be set for each propagation.

The choice of bead mass $\mu_{jk}$, originally introduced by Parrinello and Rahman,\cite{Parrinello1984} is arbitrary, with a common choice of $\mu_{jk} = m_j$ used in the literature.\cite{Ceriotti2010, Craig2004} However, a different choice can be used as a way of pre-conditioning the problem, as for a variety of Monte Carlo~\cite{Girolami2011} and optimization methods.\cite{Nocedal1999} We choose a bead mass
\begin{equation}
\mu_{jk} = m_j \frac{\omega_k^2}{\omega_N^2} \tau_0^2,
\label{beadmass}
\end{equation}
such that the frequencies of the polymer mode propagation are scaled to the constant $\tau_0^{-1}$ for all modes. This choice of mass corresponds to a multi-step integration scheme, and results in a friction coefficient $\gamma_{jk} = 2 \gamma_0 /\tau_0$. The choice of $\tau_0$ sets the time step differently between the polymer modes and the potential propagation, but also tunes the effective temperature of the momentum distribution. For the Langevin thermostat, $\tau_0$ and $\gamma_0$ set the time scale for the correlations during the stochastic integration, although $\gamma_0$ has no impact on the rest of the propagation. The tuning of $\tau_0$ and $\gamma_0$ must be done carefully, as setting the values inappropriately may lead to poor phase-space sampling (i.e. poor ergodicity can skew the results). We describe the procedure for setting the parameters in the next section.

By repeating calculations of each ensemble average, we can also obtain an estimate for the statistical uncertainty in the final tunneling splitting. The statistical uncertainties include two effects: the convergence of the ensemble average with time, and the ergodicity of the MD trajectory. If each ensemble average has an uncertainty dominated by the time convergence, then the uncertainty is simply given by the standard deviation of the points being averaged over, which gives an estimate on how reproducible the mean value is. If, however, the uncertainty is dominated by the sampling efficiency, then it is given by the standard error (the standard deviation divided by the square root of the number of repetitions), because this is the quantity that informs us about the repeatability of the mean value. We find that, typically, the Andersen thermostat tends to be dominated by statistical uncertainties in the time convergence, whereas the Langevin thermostat is dominated by the uncertainties due to ergodicity. The Andersen thermostat is a highly disruptive thermostat, which randomizes the momenta and thermalizes the trajectory repeatedly, hence the variability of the mean is dominated by the convergence of the thermalized trajectories. In contrast, the Langevin thermostat has a weaker effect on the trajectory. Hence, some regions of phase space are much less probable and the phase space is sampled less efficiently. By optimizing the value of $\gamma_0$, the Langevin thermostat allows a very efficient thermalization of each trajectory, but each trajectory may not have sampled the most representative regions of phase space.

For the computational tests below, we quote the standard error for the uncertainty of calculations performed using the Langevin thermostat, and the standard deviation for the uncertainty of calculations performed using the Andersen thermostat. These uncertainties are estimates, and only represent the \emph{statistical} uncertainties; no systematic uncertainties are quantified. The statistical uncertainties for our results typically dominate the total uncertainty, although the uncertainty from the numerical quadrature can occasionally be comparable.

\subsection{Instanton-based thermodynamic-integration path}
To maximise the efficiency of the thermodynamic integration, we need to choose an integration path that is close to the pathway minimising the free energy as a function of $\lambda$. In practice, this means we need to identify a reasonably low free energy path along which to `pull' the last bead on the polymer string, starting at a geometry ${\bf r}$ in one well and finishing at the geometry $\hat P{\bf r}$ in the symmetry-related well. For a system such as malonaldehyde, one can construct a reasonable path by inspection (in this case, pulling the hydrogen atom along a line connecting the two oxygen atoms). However, for a more complex process, such as the rearrangement dynamics in water clusters, this {\em ad hoc} approach is unlikely to work.

Here, we propose the use of the instanton `kink' connecting the symmetry-related wells. This path (often referred to as just the `instanton') is the local minimum on the linear-polymer potential surface of \eqref{un}, which is the geometry about which one evaluates harmonic fluctuations in the RPI method.\cite{Richardson2011a,Richardson2011b} The beads along the kink lie at equally spaced intervals in imaginary time on a classical trajectory along the inverted molecular potential surface, connecting the two wells in a total imaginary time $\beta\hbar$ (and one takes the limit $\beta\hbar\to\infty$).  We can be reasonably sure of following a low free energy thermodynamic integration path if, as $\lambda$ increases from $0$ to $1$, we `pull' the last bead of the polymer along the instanton kink.

The instanton can easily be found starting with the minimum energy path on the molecular potential~\cite{Richardson2011a} (calculated using the OPTIM program~\cite{OPTIM}), then performing an L-BFGS minimization~\cite{Byrd1995} of $U_N$. The value of $\lambda$ is then taken to be the integrated path length (the sum of the Euclidean distances between the points along the path), divided by the total path length. One subtlety is that we must be careful to include all the possible degenerate paths. Thus, we replace $I_i$ in \eqref{eta} by $I_i \rightarrow g_i I_i$, where $g_i$ is the number of degenerate instanton paths that result in the same PI symmetry operation.

\section{Numerical tests}
\label{modelsection}
\subsection{One-dimensional models}
We first test the PIMD method on two one-dimensional models, which were introduced by Matyus \emph{et al}.~\cite{Matyus2016a}
These models are double-well potentials of the form
\begin{equation}
V(x) = V_0 \left( \frac{x^2}{x_0^2} -1 \right)^2,
\label{doublewellpotential}
\end{equation}
with parameters $V_0=$ 1 a.u., $x_0=$ 1 a.u., $m=$ 1 a.u. (Test 1), and $V_0= 5\times 10^{-3}$ a.u., $x_0=$ 1 a.u., $m= 1822.9$~a.u. (Test 2). Test 1 corresponds to a system with a wavefunction delocalized over the two wells, which is a difficult case to treat for RPI because of the large tunneling splitting compared to the separation of the ground state and the second excited state. Test 2 corresponds to a system with a wavefunction localized in each well. RPI gives a better result for Test 2, but represents a more difficult case for the PIMD method, as the wavefunction is more localized and effectively traps the polymer in certain configurations, leading to less efficient sampling of phase space.

\begin{figure}[tb]
\centering
\includegraphics{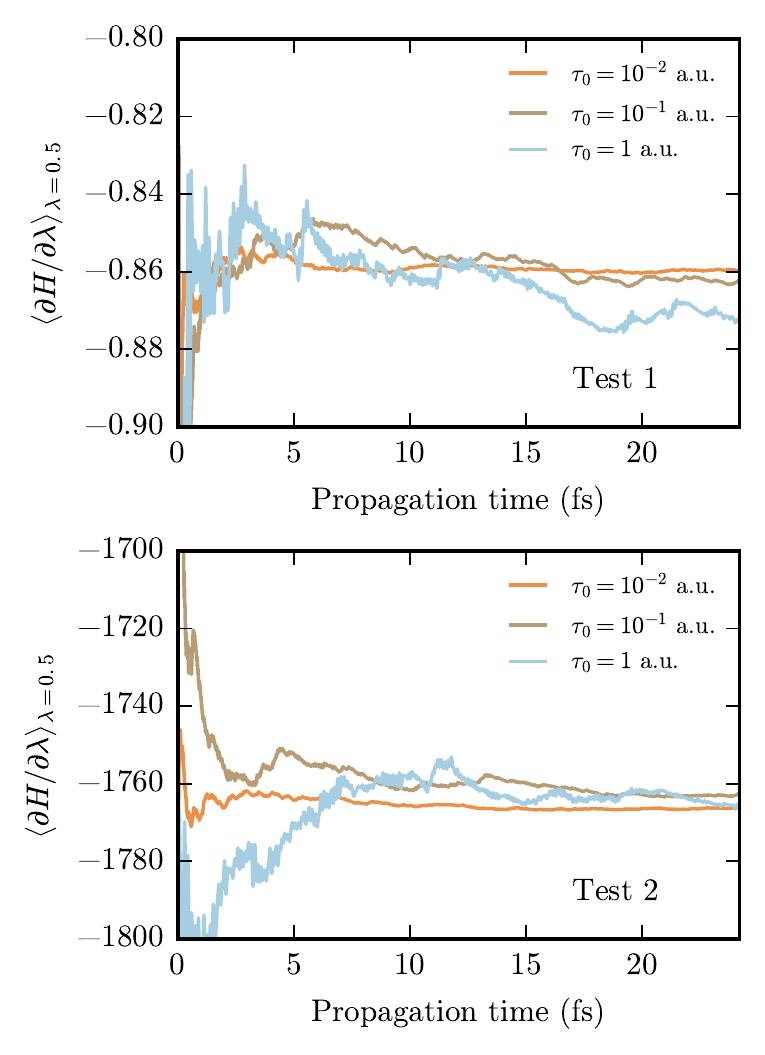}
\caption{\label{tau} Ensemble average of $\langle \partial H/ \partial \lambda \rangle$ for $\lambda=0.5$ as a function of time, calculated using the Anderson thermostat, for different values of the time constant $\tau_0$.}
\end{figure}

\begin{figure}[tb]
\centering
\includegraphics{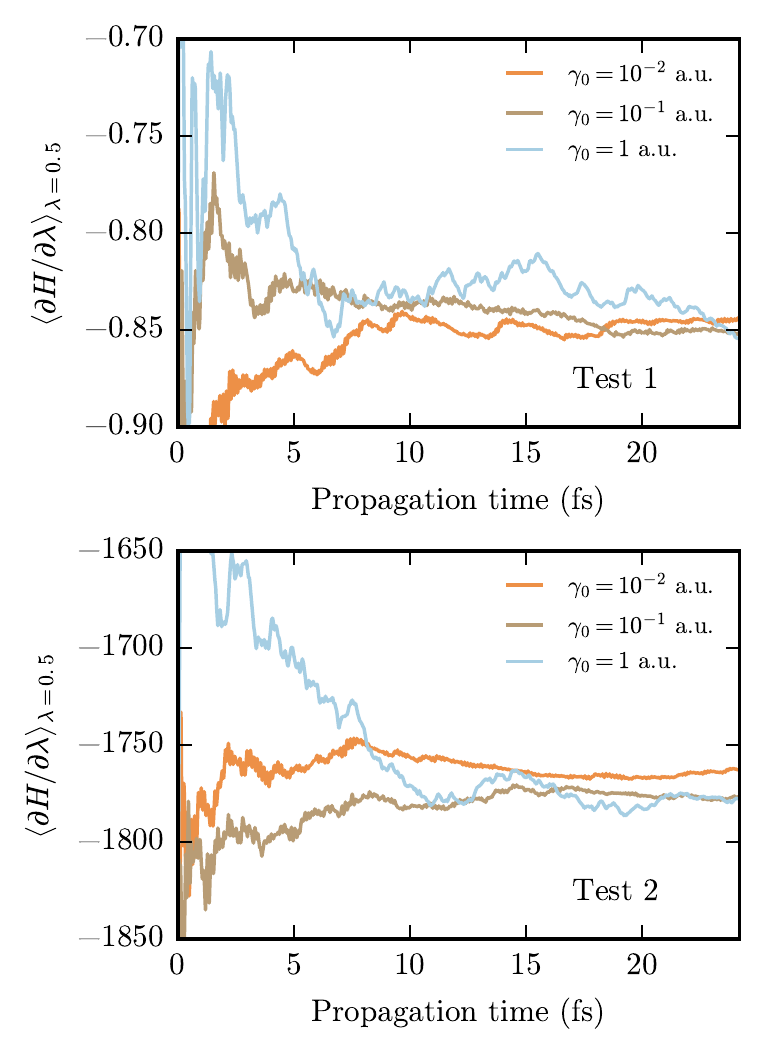}
\caption{\label{gamma} Ensemble average of $\langle \partial H/ \partial \lambda \rangle$ for Test 2 and $\lambda=0.5$, as a function of time, calculated using the Langevin thermostat. Different values of the friction constant $\gamma_0$ are shown for a reduced propagation, with a step size of $0.01$ a.u..}
\end{figure}

We first examine the effects of $\tau_0$ and $\gamma_0$. We set $\lambda=0.5$, since this corresponds to the end bead being fixed at the tunnelling barrier, which is the most difficult integral to converge.
Figures~\ref{tau} and \ref{gamma} show the ensemble average from \eqref{hamiltonianderivative} for a single propagation at different values of $\tau_0$ (for the Andersen thermostat) and $\gamma_0$ (for the Langevin thermostat), respectively. For both tests with the Andersen thermostat, the smallest value of $\tau_0$ leads to the most favourable convergence for a given choice of time step ($\Delta t= 10^{-2}$ or $10^{-3}$ a.u. in these runs). When too small a value of $\tau_0$ is chosen, either the time steps for the potential propagation become too large and an accummulation of errors prevents a successful MD run, or the run is successful but the sampling becomes biased, leading to a wrong final result.

For $\gamma_0$ the situation is similar, as shown in Fig. \ref{gamma}. For a large value of $\gamma_0$ we observe very large fluctuations, and smaller values lead to faster convergence but poor sampling. We find that the Langevin thermostat can become trapped around certain values for long periods of integration time, with occasional jumps, indicating a poor sampling efficiency. When the value of $\gamma_0$ is taken to be too low, we also observe biasing of the phase-space sampling leading to erroneous results.

In general, the Andersen thermostat transfers a broad distribution of momentum to the motion of the beads, which leads to more ergodic sampling of phase space, but slower convergence. The Langevin thermostat, in contrast, is a weaker perturbation to the motion, leading to faster convergence, but at the expense of sampling efficiency, where the stochastic perturbations are insufficient to overcome barriers between regions of phase space. The values of $\gamma_0$ and $\tau_0$ must therefore be chosen carefully to balance fast convergence with efficient sampling in both cases. Averaging over several MD trajectories can also help to reduce problems with sampling efficiency.

\begin{table*}[htb]
\footnotesize
\caption{\label{1dresults} Results of the one-dimensional model calculations using PIMD with the Langevin thermostat (L-PIMD) and the Andersen thermostat (A-PIMD), compared with variational results.\cite{Balsa1983} The parameters used for each test are listed, with $\tau_0$ given for A-PIMD and $\gamma_0$ given for L-PIMD. For all calculations, 10 integration points were used with a time step of $10^{-2}$ a.u., with each point repeated 5 times. The values for $\Delta$ and $\bar{\beta}$ are calculated from \eqref{simultaneoustanh}. For Test 1, $\beta_1=3$ a.u. and $\beta_2=4$ a.u.; for Test 2, $\beta_1=2500$ and $\beta_2=3000$ a.u..}
\begin{ruledtabular}
\begin{tabular}{lllcccc}
& ($N$, $t$, $\tau_0$ or $\gamma_0$)& Method & $I_1 (\beta_1)$ &  $I_2 (\beta_2)$ & $\Delta$ (a.u.)& $\bar{\beta}$ (a.u.)\\
\hline
Test 1 &(50, $10^3$, 0.03)& L-PIMD  & 0.71 (1) & 0.86 (1) & 0.81 (9) & 0.8 (3) \\
&(50, $10^4$, 0.03)& L-PIMD  & 0.717 (4) & 0.861 (3) & 0.79 (3) & 0.7 (1) \\
 &(50, $10^4$, 1)& A-PIMD  & 0.71 (1) &  0.86 (2) & 0.8 (1) & 0.8 (4)\\
& & Variational & 0.7183 & 0.8617  & 0.792 & 0.716\\
\hline
Test 2  & (50, $5\times 10^4$, 0.02)&L-PIMD & 6.43 (4) $\times 10^{-2}$ &8.57 (8) $\times 10^{-2}$ & 8.6 (4) $\times 10^{-5}$ & 1.00 (7) $\times 10^{3}$ \\
  & (50, $2\times 10^5$, 0.02)&L-PIMD &  6.44 (3) $\times 10^{-2}$ &8.58 (5) $\times 10^{-2}$ & 8.6 (2) $\times 10^{-5}$ & 1.0 (1) $\times 10^{3}$ \\
  & (100, $2\times 10^5$, 0.02)&L-PIMD & 6.51 (3) $\times 10^{-2}$ &8.53 (4) $\times 10^{-2}$ & 8.2 (2) $\times 10^{-5}$ & 9 (1) $\times 10^{2}$ \\
&(50, $2\times 10^5$, 1)& A-PIMD & 6.46 (6) $\times 10^{-2}$ & 8.47 (8) $\times 10^{-2}$ & 8.1 (4) $\times 10^{-5}$ & 9.0 (9) $\times 10^{2}$ \\
&(100, $2\times 10^5$, 1)& A-PIMD & 6.47 (4) $\times 10^{-2}$ & 8.6 (1) $\times 10^{-2}$ & 8.6 (5) $\times 10^{-5}$ & 9.8 (9) $\times 10^{2}$ \\
&(100, $2\times 10^5$, 0.05)& A-PIMD & 6.91 (3) $\times 10^{-2}$ &9.30 (4) $\times 10^{-2}$ & 9.6 (2) $\times 10^{-5}$ & 1.06 (3) $\times 10^{3}$ \\
& & Variational & 6.504 $\times 10^{-2}$ & 8.559 $\times 10^{-2}$  & $8.267 \times 10^{-5}$ & 924 \\
\end{tabular}
\end{ruledtabular}
\end{table*}

PIMD simulations were run for both 1D tests to calculate the tunneling splittings. The results are summarized in Table~\ref{1dresults}, along with the parameters used for the calculations. We note that an even number of integration points avoids the centre of the barrier which is associated with the largest uncertainty. Generally, the Langevin thermostat allows for a much more precise calculation of the tunneling splittings for both systems, with typically less than half the uncertainty of the Andersen thermostat. The uncertainty of each MD run scales as $t^{-1/2}$, hence an improvement of a factor of 2 in the uncertainty corresponds to a factor of 4 in terms of sampling efficiency. All the values of $I(\beta)$ agree with the variational calculations to within the statistical uncertainty, although we note that other systematic uncertainties exist that are unaccounted for, including uncertainty in the approximated numerical integral and convergence with respect to $N$. Importantly, however, the calculations remain accurate over a large range of tunneling splittings, unlike RPI, which is limited in accuracy to small tunneling splittings.\cite{Richardson2011a} However, we stress that choosing values of $\tau_0$ and $\gamma_0$ that are too small can produce a large inconsistency between the variational value of $I(\beta)$ and the PIMD calculations for small $\beta$, as seen in Table~\ref{1dresults}.

Note that throughout this article we use Eq.~\eqref{simultaneoustanh}, which requires that we compute $I(\beta)$ at just two values of $\beta$.   It is best to take the smallest reasonable value of $\beta$ to minimise the computational cost. Typically, higher values of $\beta$ require more beads, smaller time steps and longer integration time, which leads to more expensive statistics. In principle, rather than using Eq.~\eqref{simultaneoustanh}, the values of $I(\beta)$ could be fitted to Eq.~\eqref{tanhratio} using a least-squares approach, but this would require more values of $I(\beta)$ to be calculated. It is generally preferrable to limit the calculations of $I(\beta)$ to two small values of $\beta$ and improve the convergence, as this remains a less computationally intensive method of reducing uncertainties than using least-squares fitting.

\subsection{Malonaldehyde}
\begin{figure}
\centering
\includegraphics[width=3.25in]{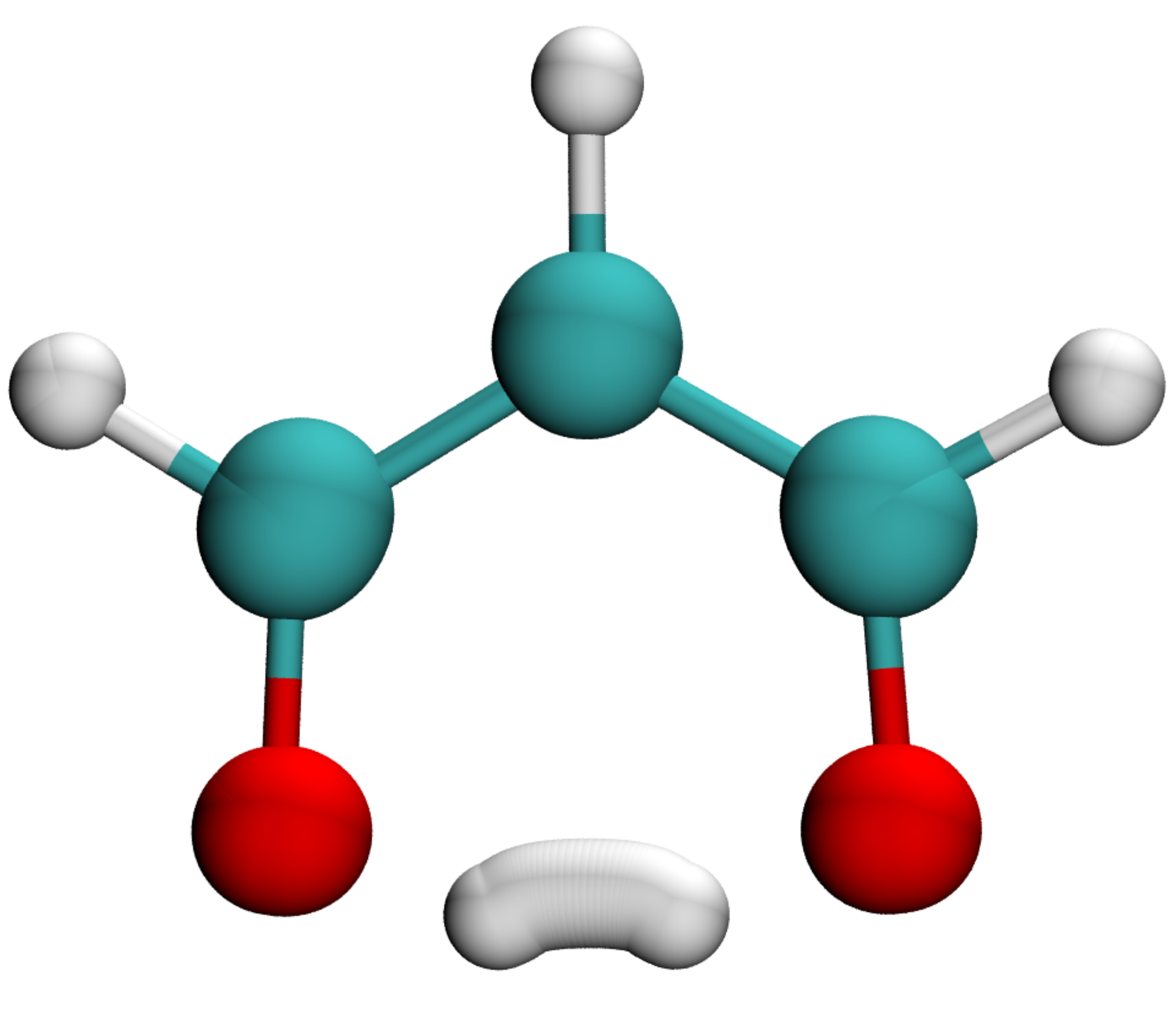}
\caption{\label{malonaldehydeinstanton} Instanton kink for proton transfer in malonaldehyde.}
\end{figure}

We next apply the PIMD method to proton transfer in malonaldehyde. We use a high accuracy potential energy surface (PES),\cite{Mizukami2014} for which DMC calculations are also available, as well as previous PIMD calculations.\cite{Matyus2016a} We include all  27 degrees of freedom.  The instanton kink for the proton transfer is shown in Fig. \ref{malonaldehydeinstanton}. 

\begin{table*}[htb]

\caption{\label{malonaldehyde} PIMD results for malonaldehyde calculated using the Langevin thermostat (L-PIMD), compared with previous results\cite{Matyus2016a} calculated using an Andersen thermostat (A-PIMD). The parameters used for each test are described in the main text. }
\begin{ruledtabular}
\begin{tabular}{lccccccc}
 $\beta$ (a.u.)&2000 & & 3000 & & 3500 & & 4000\\
\hline
\multicolumn{8}{c}{L-PIMD} \\
\hline
$100 \times I (\beta)$& 4.73 (5) & & 9.1 (1) & & 11.4 (1) & & 13.4 (1) \\
 $\Delta$ (cm$^{-1}$)& & 19.4 (4) & & 20 (2) & & 18 (2)  \\
 $\bar{\beta}$ ($\times$ 100 a.u.) & & 9.3 (3) & & 10 (2) & & 6 (3)  \\
\hline
\multicolumn{8}{c}{A-PIMD~\cite{Matyus2016a}\footnote{Uncertainties are calculated as the standard deviations of two points, and so only represent a rough estimate of the actual error.}}\\
\hline
$100 \times I (\beta)$& ---& & 8.30 (5) & & 10.4 (3)&  & 12.7 (2)  \\
$\Delta$ (cm$^{-1}$) & &--- & & 19 (3) & & 20 (3)  \\
$\bar{\beta}$ ($\times$ 100 a.u.) & & --- & & 10 (2) & & 13 (4) \\
\end{tabular}
\end{ruledtabular}
\end{table*}

\begin{table}[tb]
\caption{\label{malonaldehydesummary} Malonaldehyde tunneling splittings and, where relevant, values of $\bar{\beta}$, obtained using different methods. The computational results were performed on the same PES~\cite{Mizukami2014}, and the PIMD values are calculated using least-squares fitting of the numbers quoted in Table \ref{malonaldehyde}.}
\begin{ruledtabular}
\begin{tabular}{lcc}
Method & $\Delta$ (cm$^{-1}$) & $\bar{\beta}$ (100 $\times$ a.u.)\\
\hline
L-PIMD & 19.3 (2) & 9.2 (2)\\
A-PIMD & 19.7 (3) & 11.5 (3)\\
Instanton & 19 & \\
DMC~\cite{Mizukami2014} & 21.0 (4)& \\
Exp.~\cite{Baba1999} & 21.583\footnote{The experimental uncertainty, $6.3 \times 10^{-7}$ cm$^{-1}$, is much smaller than the number of quoted significant figures.} & \\
\end{tabular}
\end{ruledtabular}
\end{table}

The results for the malonaldehyde calculations are shown in Table~\ref{malonaldehyde}. These calculations used a step size $\Delta t=0.01$ a.u., a total propagation time $t_\mathrm{tot}= 10^5$ a.u.\ and 10 integration points, each point being repeated five times. For $\beta< 3500$ a.u., only 100 beads were required; 150 beads were needed for larger values of $\beta$. For comparison, the results from previous calculations are also quoted~\cite{Matyus2016a} (reanalyzed to ensure consistency with the L-PIMD results of this work), where the same step size and bead number were used. The improved L-PIMD algorithm runs with $10^{11}$ steps and achieves the same precision as the DMC calculations ($\sim 0.4 \; \mathrm{cm}^{-1}$), whereas the re-analyzed A-PIMD from the previous work ran with $3\times 10^{11}$ potential calls. These numbers should be compared with the DMC calculations, which took $5.5 \times 10^9$ function calls.\cite{Mizukami2014}

The final calculated splittings in Table~\ref{malonaldehydesummary} (which were obtained using least-squares fitting of all the results in Table~\ref{malonaldehyde}) show a discrepancy of about $10\%$ between the PIMD and DMC results. The L-PIMD results from this work were tested for convergence in the number of quadrature points used, the values of $\beta$, the number of beads, the value of $\gamma_0$, and the value of $\tau_0$; all results were within the quoted statistical uncertainty. The remaining discrepancy remains unexplained. One possibility is an underestimated contribution to the uncertainty of DMC due to the fixed-node approximation. However, there is no evidence in Ref.~\onlinecite{Mizukami2014} of such a large uncertainty, and an investigation into the uncertainties arising in the DMC calculations is beyond the scope of this article. Another possible explanation is that the PIMD result could be contaminated by tunnelling splittings from excited rotational states (and we return to this possible source of error in Secs.~\ref{watersection} and \ref{discussionsection}).

\subsection{Water Dimer}
\label{watersection}
\begin{figure}
\centering
\includegraphics[width=2.75in]{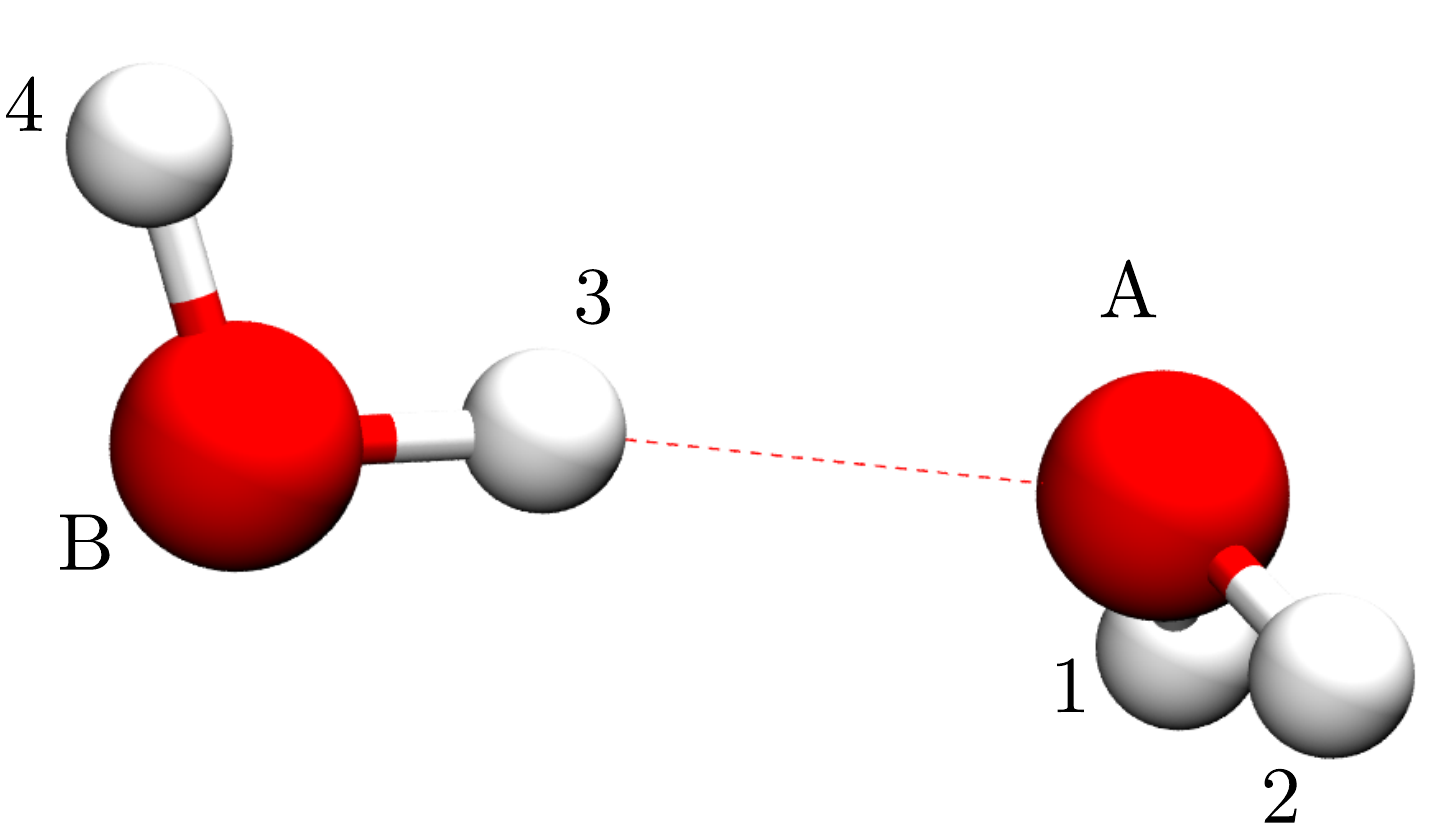}
\caption{\label{waterdimer} Global-minimum structure of the water dimer on the MB-pol PES.\cite{Babin2013} The oxygen atoms are labelled A and B, and the hydrogen atoms 1-4, following the convention of Ref.~\onlinecite{Dyke1977}.}
\end{figure}

In the absence of covalent bond breaking, water dimer has eight symmetry-related global minima,\cite{Dyke1977} resembling the `donor-acceptor' structure of Fig.~\ref{waterdimer}. \footnote{The PI symmetry gives 16 minima, but half of these may be rotated without tunnelling into the other half, on account of the plane of symmetry\cite{Dyke1977}} It is well known that there are five distinct rearrangement mechanisms connecting these minima. Following previous work,\cite{Hougen1985,Richardson2011b} we label these the acceptor (A), donor (D), geared (G), anti-geared (AG) and bifurcation (B) tunneling pathways. Further details of the symmetry analysis of the water dimer used in our calculations are given in Appendix~\ref{watersymmsection}.

The first step in the calculation is to define the integration paths between the wells connected by the five tunnelling pathways. As explained in Sec.~IIIB, we choose these paths to follow the instanton kinks connecting the wells. Following previous RPI calculations on the dimer,\cite{Richardson2011b} the minimum energy paths between the wells were used as initial guesses to find the instanton kinks with a standard L-BFGS algorithm.\cite{Byrd1995} The minimum energy paths were calculated using the OPTIM program,\cite{OPTIM} modified to use the MB-pol water PES~\cite{Babin2013,Babin2014} (which is the PES that we use for all the water dimer calculations).

\begin{figure}
\centering
\includegraphics{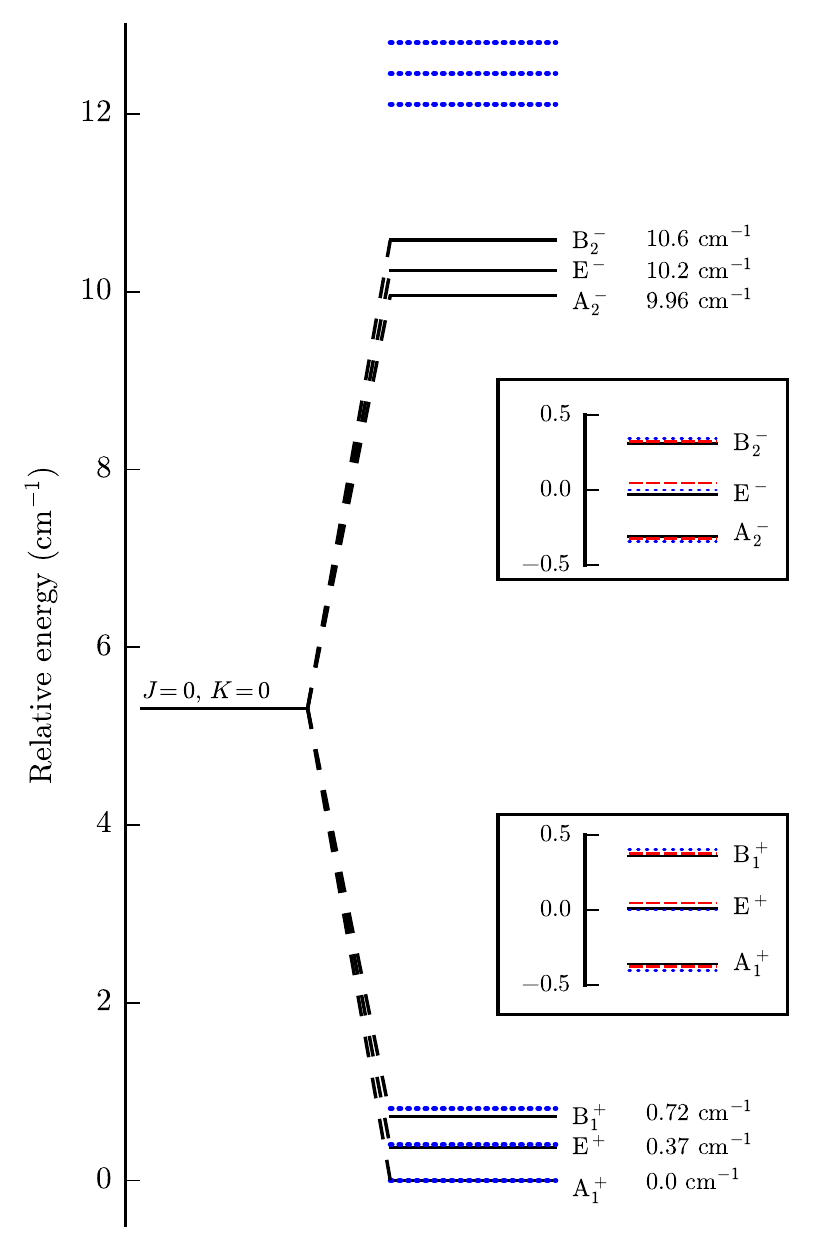}
\caption{\label{levels} Energy levels for the ground rotational state of the water dimer. The PIMD results (black solid lines) are compared with variational results taken from Ref.~\onlinecite{Babin2013} (blue dotted lines). The insets compare the PIMD, variational and experimental\cite{Zwart1991} (red dashed lines) results for the corresponding upper and lower branches, relative to the branch centers.~\cite{Note2}}
\end{figure}

The splitting pattern obtained from the PIMD calculations is summarised in Fig. \ref{levels}, where it is compared with variational and experimental results.\cite{Wang2018, Groenenboom2000, Note2} The pattern was obtained from the individual density matrix ratios $I_i$ for the five tunnelling pathways, given in Table~\ref{watertable}. The acceptor and geared (also known as the donor-acceptor interchange) motions contribute the most to the splitting pattern. For  these two paths, we used 200 beads with a time step of $10^{-2}$ a.u.\ for a total integration time of $5 \times 10^{4}$ a.u. (a total of $5 \times 10^6$ steps). The parameters $\tau_0$ and $\gamma_0$ were both set to unity, to avoid any problems with sampling the full phase space; such sampling problems were observed for small values ($< 10^{-1}$) of these parameters. For the other three paths, we used 100 beads and integrated for $2 \times 10^4$ a.u. with $\gamma_0=0.5$. For all paths, 16 Gauss-Legendre integration points were used with each point repeated 5 times, with calculations spread over 16 2.6 GHz processors.  The acceptor and geared calculations took 200 hours to complete, significantly less than state-of-the-art variational calculations.\cite{Wang2018}

\begin{table}[htb]
\caption{\label{waterresults} Density matrix ratios $I_i$, calculated by PIMD (using the parameters given in the text) for the five tunnelling pathways of water dimer. Each $I_i$ was calculated at two values of $\beta$, allowing $\Delta$ and $\bar{\beta}$ to be obtained by combining the expressions for $I_\nu$ in Table~\ref{watertable} with \eqref{simultaneoustanh}.}
\begin{ruledtabular}
\begin{tabular}{lcc}
$I_i$ &  $\beta=12,000$ a.u. & $\beta=20,000$ a.u. \\
\hline
$I_\mathrm{A}$ & 1.34 (8) $\times 10^{-2}$  & 1.02 (8) $\times 10^{-1}$ \\
$I_\mathrm{G}$ & 9 (1) $\times 10^{-3}$ & 1.4 (2) $\times 10^{-2}$ \\
$I_\mathrm{AG}$ & 7.7 (3) $\times 10^{-4}$ & 3.3 (2) $\times 10^{-3}$ \\
$I_\mathrm{D}$ & 1.9 (1) $\times 10^{-5}$ & 3.7 (4) $\times 10^{-4}$ \\
$I_\mathrm{B}$ & 1.06 (7) $\times 10^{-6}$ & 4.8 (7) $\times 10^{-6}$ \\
\end{tabular}
\end{ruledtabular}
\end{table}

 From Fig. \ref{levels}, the PIMD estimate of the acceptor splitting is $(E_{\mathrm{A}_2^-} + E_{\mathrm{B}_2^-})/2-(E_{\mathrm{A}_1^+} + E_{\mathrm{B}_1^+})/2= 9.9 \pm 1.2 \, \mathrm{cm}^{-1}$, which is about 19$\%$ smaller than the variational result of 12.05 $\mathrm{cm}^{-1}$. This discrepancy is almost certainly the result of errors in the PIMD calculations, as the variational calculations used the same MB-pol PES,\cite{Babin2013} and the only approximation they employed (an adiabatic separation) has been shown in recent full-dimensional calculations\cite{Wang2018} (on a different PES) to affect the acceptor splitting by only about 0.2 cm$^{-1}$. Given that the acceptor splitting depends strongly on $K$, such that the $K=1$ splitting is about a quarter of the $K=0$ splitting,\cite{Groenenboom2000} the most likely cause of the PIMD error is contamination of $I_{A}$ from nearby rotational levels (although one can never rule out in a PIMD calculation the possibility of sampling errors from non-ergodicity). Clearly the PIMD acceptor splitting is better than the RPI splitting (calculated previously in Ref.~\onlinecite{Richardson2011b}), but is not good enough for quantitative comparison with experiment. 
 
The differences between the PIMD and variational results for the interchange splittings (dominated by the geared and anti-geared paths) are much smaller, and indeed the upper branch splitting $E_{\mathrm{B}_2^-}- E_{\mathrm{A}_2^-}= 0.62 \pm 0.04 \, \mathrm{cm}^{-1}$ agrees with the variational result to within the combined statistical uncertainty of the two calculations (including an estimate of the uncertainty associated with the adiabatic approximation). The lower interchange splitting lies slightly outside the uncertainty, with $E_{\mathrm{B}_1^+} - E_{\mathrm{A}_1^+} = 0.72 \pm 0.04 \, \mathrm{cm}^{-1}$. The energies of the degenerate E$^\pm$ states are slightly shifted by the donor tunneling motion, which is consistent with previous observations.\cite{Zwart1991}

\section{Conclusions}
\label{discussionsection}

We have improved the PIMD methodology of Ref.~\onlinecite{Matyus2016a} using a Langevin thermostat (which gives a modest gain in efficiency) and by introducing an integration path based on the instanton kink. These improvements (especially the latter) have allowed us to carry out PIMD simulations on water dimer, which is an important multi-well tunnelling system that gives rise to a large splitting (comparable to the rotational spacings) and small splittings (much smaller than the rotational spacings).

We find that the size of the splittings plays a key role in the success of the PIMD method. For small splittings (in water dimer) the PIMD results agreed with accurate variational results to within (or slightly outside) the statistical uncertainty in the sampling ($\sim10\%$). For the large (acceptor) splitting the results disagreed by about $20\%$ with the variational results. This error is too big to be attributed entirely to statistical uncertainty, and is most likely the result of contamination from neighbouring rotational states;  a similar but smaller error, probably attributable to the same effect, was found in the PIMD results for malonaldehyde. In future work, it may be possible to use the ideas of Ref.~\onlinecite{Matyus2016b} to project out these excited states, and thus converge to accurate PIMD results for large splittings; however, this is likely to require much more expensive sampling.

We conclude that PIMD is  the stochastic method of choice for calculating splittings that are much smaller than the rotational energy spacings. Such splittings are extremely difficult to calculate by DMC (which, in contrast, is certainly the method of choice for large splittings, provided one has prior knowledge of the nodal surface).  However, it may well turn out that stochastic methods are not needed to calculate these small splittings, given the success of the RPI method, which has been found to work well for small splittings in water dimer and formic acid dimer, but fails for large splittings owing to the `floppiness' of the polymers. Future tests of the PIMD method on the water hexamer, which has small splittings, where RPI results disagree by a factor of two with experiment\cite{Richardson2016}, may answer this question.

The software used to create the data in this article is available at https://doi.org/10.17863/CAM.20999.

\acknowledgments{We are grateful to E.\ M\'atyus for sharing her programs at initial stages of the project, and to J.\ O.\ Richardson for useful discussions. The project was funded by the United Kingdom Engineering and Physical Sciences Research Council (EPSRC).}

\appendix
\section{Autocorrelation Time}
\label{autocorrelationsection}
To derive the autocorrelation time, there are a number of steps we need to take. Firstly, our probability distributions are all Gaussians, so we can use the property of Gaussian correlation functions~\cite{Zwanzig2001, Gardiner1985,Wang1945}
\begin{equation}
\langle x_1 x_2 x_3 x_4 \rangle = \langle x_1 x_2 \rangle \langle x_3 x_4 \rangle + \langle x_1 x_3 \rangle \langle x_2 x_4 \rangle + \langle x_1 x_4 \rangle \langle x_2 x_3 \rangle.
\label{gaussianproperty}
\end{equation}
Secondly, we will be working mainly with the correlation functions involving positions and momenta. This means that, given a position correlation function $\langle q(0) q(t) \rangle$, we can derive the other necessary correlation functions using
\begin{subequations}
\begin{align}
\langle q (0) \pi (t) \rangle &= \mu \frac{\mathrm{d}}{\mathrm{d}t} \langle q(0) q(t) \rangle\\
\langle \pi (0) \pi (t) \rangle &= \mu \frac{\mathrm{d}}{\mathrm{d}t} \langle q(0) \pi(t) \rangle.
\end{align}
\label{momentumcorrelations}%
\end{subequations}

The crucial first element in obtaining the correlation time is deriving $\langle q(0) q(t) \rangle$. To obtain this quantity we need to solve the stochastic Langevin equations. One way to solve the equations in \eqref{langevin} is by rewriting them as~\cite{Zwanzig2001}
\begin{equation}
\frac{\mathrm{d}}{\mathrm{d}t}\vec{\eta} = \hat{A}\cdot\vec{\eta} + \sqrt{\frac{2 m \gamma}{\beta_N}} \vec{\xi}(t),
\label{veclangevin}
\end{equation}
where $\vec{\eta} = (q, \pi)^\mathrm{T}$, $\vec{\xi} = (0, \xi)^\mathrm{T}$, and
\begin{equation}
\hat{A} = \begin{pmatrix}
0 & \mu^{-1}\\
-m \omega^2 & -\gamma
\end{pmatrix}.
\label{matrix}
\end{equation}
The solution to \eqref{veclangevin} is then
\begin{equation}
\vec{\eta}(t) = e^{\hat{A} t}\cdot \vec{\eta} (0) + \int^t_0 \mathrm{d}t' e^{\hat{A} (t-t')} \cdot \vec{\xi}(t').
\label{propsolution}
\end{equation}
In general, the integral on the right hand side of \eqref{propsolution} will be a vector of statistically-independent Gaussians, both with zero-mean. Thus, when we need to work out a correlation function, we will be able to neglect the stochastic term as this will average to zero.

The homogeneous, second order linear differential equation that we have is simply
\begin{equation}
\frac{\mathrm{d}^2 q}{\mathrm{d}t^2} + \gamma \frac{\mathrm{d}q}{\mathrm{d}t}  + \frac{m \omega^2}{\mu} q =0,
\label{dampedODE}
\end{equation}
with solution
\begin{equation}
\begin{split}
q (t) &= e^{-\gamma t/2} \left[ q(0) \cos(\Omega t) \right.\\
&+ \left.\frac{2 \pi(0) + \mu \gamma q(0)}{2 \mu \Omega}\sin(\Omega t)\right],
\end{split}
\label{dampedsolutions}
\end{equation}
where $\Omega^2 = m\omega^2/\mu - \gamma^2/4$. We next find an expression for $q(t) q(t')$ and take $t'=0$ to obtain
\begin{equation}
\begin{split}
q(t) q (0) &= e^{-\gamma t/2} \left[ q(0)^2 \cos(\Omega t) \right.\\
&+ \left. \frac{2 \pi(0) q(0) + \mu \gamma q(0)^2}{2 \mu \Omega}\sin(\Omega t)\right].
\end{split}
\label{qiqi}
\end{equation}
Taking the time average, and noting that $\pi(0)$ and $q (0)$ are statistically independent quantities with mean 0 such that $\langle \pi(0) q (0) \rangle = 0$, we obtain the position correlation function
\begin{equation}
\langle q (t) q (0) \rangle = \langle q (0)^2 \rangle e^{-\gamma t/2} \left[ \cos(\Omega t) + \frac{\gamma}{2\Omega} \sin(\Omega t) \right].
\label{positioncorrelation}
\end{equation}

To work out $\langle q (0)^2 \rangle$, let us first consider the mean value of the Hamiltonian,
\begin{equation}
\langle H \rangle = \frac{\langle \pi^2\rangle}{2 \mu} + \frac{1}{2} m \omega^2 \langle q^2 \rangle.
\label{meanhamiltonian}
\end{equation}
We know that at equilibrium the momenta follow a Maxwell-Boltzmann distribution, therefore $\langle \pi^2 \rangle = \mu/\beta_N$. We also know that the expectation of the energy should be $k_\mathrm{B} T = 1/\beta_N$. Combining these results we obtain
\begin{equation}
\langle q (0)^2 \rangle = \frac{1}{m \omega^2 \beta_N}.
\label{initialposition}
\end{equation}

Now that we have the position correlation function, we use \eqref{momentumcorrelations} to find the position-momentum correlation function
\begin{equation}
\langle q(0) \pi(t) \rangle = - \frac{1}{\beta_N \Omega} e^{-\gamma t/2} \sin(\Omega t),
\label{positionmomentum}
\end{equation}
and the momentum correlation function
\begin{equation}
\langle \pi (t) \pi (0) \rangle = \frac{\mu}{\beta_N} e^{-\gamma t/2} \left[\frac{\gamma}{2\Omega} \sin(\Omega t) - \cos(\Omega t) \right].
\label{momentumcorrelation}
\end{equation}

Using the Hamiltonian in \eqref{polymerhamiltonian}, we expand the correlation function $\langle H(0) H(t) \rangle$, retaining only the time-dependent terms, and making use of the Gaussian property in \eqref{gaussianproperty} to obtain
\begin{equation}
\begin{split}
\langle \delta H(0) \delta H(t) \rangle &= \frac{1}{2 \mu^2} \langle \pi (0) \pi (t) \rangle^2 + \frac{m \omega^2}{\mu} \langle \pi (0) q (t) \rangle^2\\
&+ \frac{1}{2} m^2 \omega^4 \langle q (0) q (t) \rangle^2.
\end{split}
\label{expandedhamiltoniancorrelation}
\end{equation}
It then only requires combination of expressions \eqref{positioncorrelation}, \eqref{initialposition}, \eqref{positionmomentum} and \eqref{momentumcorrelation}, before integrating over time. The final integrand reads
\begin{equation}
\begin{split}
\langle \delta H(0) \delta H(t) \rangle &=  \frac{e^{-\gamma t}}{\beta_N^2} \left[ \cos^2(\Omega t) + \frac{\gamma^2}{2 \Omega^2} \sin^2(\Omega t) \right. \\
  &+ \left. \frac{m \omega^2}{\mu \Omega} \sin^2(\Omega t) \right].
\end{split}
\label{integrand}
\end{equation}
We use the standard integrals~\cite{Gradshteyn2007}
\begin{subequations}
\begin{align}
\int^\infty_0 \mathrm{d}t e^{-\gamma t} \cos^2 (\Omega t) &= \frac{2\Omega^2 + \gamma^2}{\gamma(\gamma^2 + 4\Omega^2)}\\
\int^\infty_0 \mathrm{d}t e^{-\gamma t} \sin^2 (\Omega t) &= \frac{2\Omega^2}{\gamma(\gamma^2 + 4\Omega^2)},
\end{align}
\label{integrals}%
\end{subequations}
and divide by the normalization $\langle H^2 \rangle - \langle H \rangle^2 = \beta_N^{-2}$ to obtain the final result in \eqref{finalautocorrelation}.

\section{Symmetry Analysis of Water Dimer}
\label{watersymmsection}
Water dimer has previously been extensively studied, and as such its symmetry properties are well known.\cite{Dyke1977} The character table for the relevant motions is shown in Table~\ref{watertable}, which summarizes the important symmetry considerations needed for the PIMD calculations. There are 12 irreps in the $\mathrm{G}_{16}$ PI group (isomorphic with the $D_{4h}$ point group~\cite{Bunker1998}), but the $\sigma_h$ plane of symmetry means that these can be factorized into two groups of 6 levels. Assuming the vibrational and rotational motions can be factorized (an assumption routinely used in calculations of tunneling splittings~\cite{Richardson2011b, Matyus2016a, Richardson2011a}), the ground rotational state splits into six levels: $\mathrm{A}_1^+$, $\mathrm{E}^+$, $\mathrm{B}_1^+$, $\mathrm{A}_2^-$, $\mathrm{E}^-$, and $\mathrm{B}_2^-$. As there are only six non-degenerate wells, only two sets of five PIMD calculations need be performed to connect the initial well to the other wells. The connection between the symmetry operations of the $\mathrm{G}_{16}$ PI group and the wells (labelled as in the main text) is also given in Table~\ref{watertable}, along with the formulae for the various values of $I_\nu$ (where $\nu$ is a symmetry label corresponding to an energy level) in terms of the values of $I_i$ (where $i$ labels a well).

\begin{table*}[htb]
\footnotesize
\caption{\label{watertable} Character table for the PI group of water showing the relevant tunneling motions, the name of the tunneling motion, the corresponding PI operation, and the resulting expression for $I_\nu$ in terms of the various $I_i= \rho(\vec{r}, \hat{P}_i \vec{r}, \beta)/\rho(\vec{r}, \vec{r}, \beta)$ as obtained from \eqref{multiwelltanh}. For the PI operations, brackets with two labels indicate a permutation of the two atoms, and brackets with four labels indicate a cyclic permutation~\cite{Dyke1977}. A degeneracy factor of $g=2$ is used for the acceptor path, and $g=4$ for the bifurcation path.}
\begin{ruledtabular}
\begin{tabular}{cccccccc}
Vibrational level & \multicolumn{6}{c}{Characters of PI Operations} & $I_\nu$\\
\hline
 & I & A & D & AG & G & B & \\
 & E & (34) & (12)       & (AB)(13)(24)  & (AB) (1324)     & (12)(34) & \\
 & (12)*    & (12)(34)*  &  E*  & (AB)(14)(23)  & (AB) (1423)     & (34)* & \\
 &       &            &      & (AB)(1324)*   & (AB) (13)(24)*  &  & \\
 &       &            &      & (AB)(1423)*   & (AB) (14)(23)*  &  & \\
\hline
$\mathrm{A}_1^+$ & 1 & 1 & 1  & 1 & 1 & 1 & $I_{\mathrm{A}_1^+}=0$\\
$\mathrm{E}^+$ & 1 & 1 & -1 & 0 & 0 & -1 & $I_{\mathrm{E}^+}=\frac{4 I_\mathrm{B} + I_\mathrm{D}+I_\mathrm{G} + I_\mathrm{AG}}{1 + 2 I_\mathrm{A}+I_\mathrm{G} + I_\mathrm{AG}}$\\
$\mathrm{B}_1^+$ & 1 & 1 & 1 & -1 &-1 & 1 & $I_{\mathrm{B}_1^+}=\frac{2I_\mathrm{G} + 2I_\mathrm{AG}}{1 + 2 I_\mathrm{A}+ I_\mathrm{D}+ 4 I_\mathrm{B}}$\\
$\mathrm{A}_2^-$ & 1 & -1 & -1 & -1 & 1 & 1 & $I_{\mathrm{A}_2^-}=\frac{2 I_\mathrm{A} +I_\mathrm{D} + 2I_\mathrm{AG}}{1 + 2I_\mathrm{G}+ 4 I_\mathrm{B}}$\\
$\mathrm{E}^-$ & 1 & -1 & 1 & 0 & 0 & -1 & $I_{\mathrm{E}^-}=\frac{4 I_\mathrm{B} + 2 I_\mathrm{A}+I_\mathrm{G} + I_\mathrm{AG}}{1 + I_\mathrm{D}+I_\mathrm{G} + I_\mathrm{AG}}$\\
$\mathrm{B}_2^-$ & 1 & -1 & -1 & 1 & -1 & 1 & $I_{\mathrm{B}_2^-}=\frac{2 I_\mathrm{A} +I_\mathrm{D} + 2I_\mathrm{G}}{1 + 2I_\mathrm{AG}+ 4 I_\mathrm{B}}$\\
\end{tabular}
\end{ruledtabular}
\end{table*}

\bibliography{pimd}
\end{document}